**Non-destructive characterization techniques for battery performance and lifecycle assessment**

Charlotte Gervillié-Mouravieff[1], Wurigumula Bao[2], Daniel A Steingart[3,4], Ying Shirley-Meng[1,2]*

[1]Department of NanoEngineering, University of California, San Diego, La Jolla, CA 92093, USA

[2]Pritzker School of Molecular Engineering, University of Chicago, Chicago, IL 60637, USA

[3]Department of Chemical Engineering, Columbia University, New York, NY 10027, USA

[4]Columbia Electrochemical Energy Center, Columbia University, New York, NY 10027, USA

*Corresponding author: shirleymeng@uchicago.edu

**Abstract**

As global energy demands escalate, and the use of non-renewable resources become untenable, renewable resources and electric vehicles require far better batteries to stabilize the "new energy" landscape. To maximize battery performance and lifetime, understanding and monitoring the fundamental mechanisms that govern their operation throughout their life cycle is crucial. Unfortunately, from the moment batteries are sealed until their end-of-life, they remain a "black box," and our current knowledge of a commercial battery's health status is limited to current (I), voltage (V), temperature (T), and impedance (R) measurements, at the cell or even module level during use. Electrochemical models work best when the battery is new, and as "state reckoning" drifts leading to an over-reliance on insufficient data to establish conservative safety margins resulting in the systematic under-utilization of cells and batteries. While the field of *operando* characterization is not new, the emergence of techniques capable of tracking commercial battery properties under realistic conditions has unlocked a trove of chemical, thermal, and mechanical data that has the potential to revolutionize the development and utilization strategies of both new and used lithium-ion devices.

In this review, we examine the latest advances in non-destructive *operando* characterization techniques, including electrical sensors, optical fibers, acoustic transducers, X-ray-based imaging and thermal imaging (IR camera or calorimetry), and their potential to improve our comprehension of degradation mechanisms, reduce time and cost, and enhance battery performance throughout its life cycle.

**Introduction**

Non-destructive characterization techniques illuminate and reveal the world around us, enabling crucial insights while preserving the utility of the subjects being examined. Much like non-invasive medical records extend our understanding of human health without harming the body, non-destructive characterization in batteries to provide critical data for optimizing performance and longevity, all without compromising the battery's structural integrity. Therefore, in the context of the climate emergency, where resources become increasingly scarce, the development of battery health records becomes vital[1]. Achieving this goal, however, is not without its challenges. Batteries are living objects involving dynamic electrochemical and chemical reactions, and electronic and ionic limitations. Furthermore, the presence of various parasitic reactions, including electrode-electrolyte interface instability, lithium plating, cathode and anode degradation, and electrolyte decomposition, have a considerable impact on battery performance. Hence, understanding and mitigating battery degradation mechanisms is crucial to enhance battery performance and ensuring long-term durability.

Researchers and engineers have developed several characterization techniques to better understand battery degradation. Using spectroscopic techniques (XRD, Raman, FTIR, UV-vis, NMR), microscopy techniques (SEM, TEM, AFM), thermal techniques (DSC, TGA) or gas analysis (GC, MS) they have been able to probe the fundamental aspects of batteries and optimize their behavior under different conditions[2–4]. However, most of these techniques require disassembling (or "tear-down" of) the cell for postmortem characterization, which hinders a predictive understanding of battery behavior due to the lack of real-time information, potential sample alteration, and inability to capture global and transient phenomena. Some lab scale cell designs preserved some commercial constraints while enabling characterization to be conducted in-situ with the cell in its fully assembled state[5–8]. Nevertheless, most of those lab-scale cells utilize Li metal as a counter/reference electrode in flooded electrolyte conditions, often overlooking the intricate reaction mechanisms between the electrolyte, anode, and cathode[9].

Understanding the evolution of commercial cells is imperative not only to maximize their performance and lifespan but also to evaluate their end-of-life for second-life applications or recycling purposes. However, the challenge becomes more considerable for the industry, as from the moment batteries are sealed until their end-of-life, they remain a "black box," and the current knowledge of their health status is limited to current (I), voltage (V), impedance (R) and sometimes temperature (T) measurements, at the cell or even module level during use (see Fig. 1). This lack of comprehensive knowledge leads to time-consuming trial-and-error development processes, an over-reliance on imperfect data to establish safety margins and a systematic under-utilization of cells and batteries. Consequently, battery conditioning, which includes processes like electrolyte wetting and interface formation, can take up to three weeks and contribute to 48% of the entire manufacturing cost of battery[10]. Despite this extended and intricate process, a notable portion, ranging from 5 % to 10 % of production capacity, still ends up at production scrap[11]. Another undeniable consequence is the premature disposal of batteries considered "dead" when they have lost 20% of their initial capacity while abandoning the remaining 80% as waste. Recently techniques capable of tracking commercial battery properties under realistic conditions have unlocked a trove of chemical, thermal, and mechanical data with the potential to accelerate and optimize the development and utilization strategies of lithium-ion devices, both new and used[12,13]. In this review, we will examine the latest advances in non-destructive characterization techniques and their associated benefits in understanding the life of batteries during their three main life stage: during the manufacturing process, during their usage, and finally at the end of their life (see Fig. 1). As batteries progress through these stages, the type of integration and the parameters to be monitored evolve, necessitating integration of appropriate characterizations. For the battery manufacturing process, we will explore characterizations capable of detecting manufacturing defects, observing wetting phenomena, and assessing interface formations. Turning to in use or "on-line" integration, we will provide an overview of sensors that could seamlessly integrate into electrical vehicles battery packs, presenting novel safety and performance thresholds to integrate into the battery management system (BMS), focusing on optical fiber sensing. As batteries approach their end-of-life, we will discuss the importance of characterizing this relatively unexplored state for the purposes of recycling and second-life applications. Finally, we will discuss the potential of combining these techniques to drive "battery passport" initiatives, which can improve battery data analysis and traceability and contribute to the advancement of new sustainable battery technologies (Na-ion, solid-state battery).

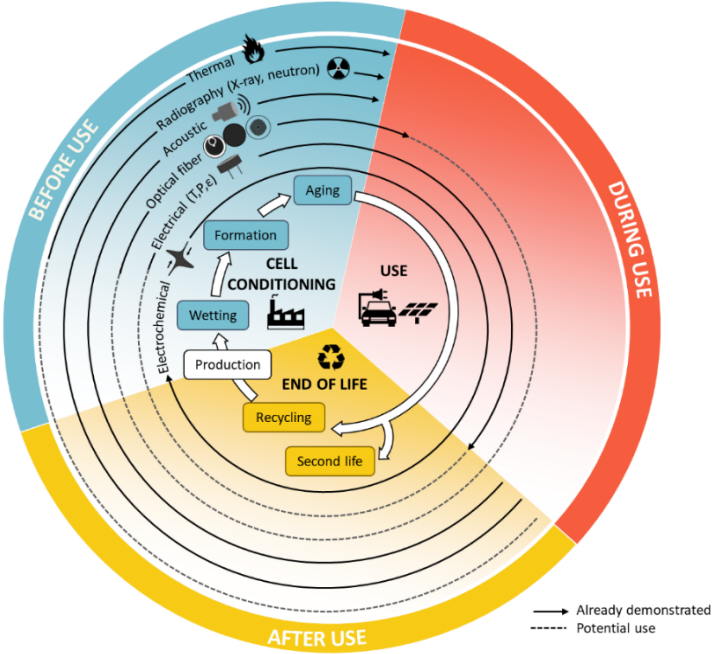

**Fig. 1 | A roadmap of non-destructive characterization techniques for comprehensive commercial battery life cycle analysis.** Non-destructive characterization being used for commercial batteries have been listed. The solid line in the figure indicates that published research has utilized the technique to characterize a particular stage of battery life: before use, during use or after use. The dashed line represents our perspective on potential future applications of the technique. The absence of a line means that the technique cannot be used in this part of the battery's life.

## I. Before use: Non-destructive characterizations to understand battery manufacturing process.

Battery manufacturing plays a crucial role in achieving optimum performance and longevity. From electrode production to cell assembly and battery electrochemistry activation, all the steps of battery production significantly impact the cell's qualities and electrochemical properties. As the manufacturing step advances, each process becomes more sensitive as it retains the value added by the previous steps. Thus, rejecting cells after the formation step results in significant time and money losses.

Before manufacturing even begins, characterizations are required to control the quality of incoming materials. Techniques like inductively coupled plasma optical emission (ICP-OES), optical microscopy or scanning electron microscopy (SEM) and energy-dispersive spectroscopy (EDS) are commonly used to help identify the impurity concentration in the materials[14]. Throughout the material production process, the QA/QC (quality assurance/ quality control) for both material and electrode are thorough. This includes morphology and structural parameters characterizations that are realized through various techniques, including SEM, XRD, profilometry or laser scanning. Additionally, the distribution of chemical species in the electrode can be examined using vibrational spectroscopies such as Fourier transform infrared (FTIT) and Raman[15,16].

However, once the cell assembly is completed, the jelly roll is enclosed within a bag or can, rendering its observation impossible. Furthermore, when the wetting is completed, batteries are hermetically closed and essentially become a 'black box,'. From this moment, quality control only relies on electrochemical performance metrics like electrochemical impedance spectroscopy (EIS), open circuit voltage (OCV), and capacity[17]. Unfortunately, these parameters only offer average values for the entire cells and lack detailed imaging and chemical data that could uncover the origins of defects or the battery's dynamic behavior. Moreover, most electrochemical techniques require sophisticated instrumentation, slow data acquisition times, and expertise in electrochemistry and data fitting techniques[18]. Consequently, even cells passing the quality control may exhibit post-production variations in capacity aging trends and safety issues[19,20]. The large number of variables during the conditioning process and limited knowledge about their aggregated effect on product properties necessitate parameterization through experimentation, which is highly time-consuming and can result in lengthy wetting and aging processes. For instance, wetting and formation alone can take up to three weeks and contribute to 48% of the entire manufacturing cost[10]. Additionally, despite the efforts makes to optimize the conditioning process, the lack of understanding of the different steps still leads to 5 % to 10 % of production capacity becoming production scrap[11]. Therefore, non-destructive characterizations are vital for enhancing efficiency, reducing costs, and minimizing scrap rates associated with battery production. In the following paragraphs, we will discuss non-destructive characterization techniques able to enhance our understanding of battery conditioning steps after battery assembly and evaluate the feasibility of these options (Fig. 2).

*Battery assembly and packaging*

Batteries are assembled in a dry room using a specific format (round wound, prismatic wound, stacked, Z-folded) and contacted internally, usually by ultrasonic or laser welding, before being inserted into a housing (hard case or pouch bag). Defects in the battery assembly process can have serious effects on battery safety and performance, such as compromised separator integrity leading to short circuits and thermal runaway, and electrode overhang causing dendrite growth and short-circuits. These defects can also hinder the battery performance resulting in degraded capacity, increased impedance, higher degradation rate, and reduced useful life.

Precise characterization of cell failures can help sort the cells and save the cost of electrolyte injection. The high-potential test in battery cell production is a traditional quality control procedure, where battery cells are subjected to high voltages to identify any separator defects or weaknesses, ensuring the battery's safety and reliability[21]. This test helps to roughly sort cells by detecting short circuits. Computed tomography (CT) is a spatially nondestructive method that combines a series of X-ray measurements taken from different angles to produces detailed cross-sectional images, allowing to quantify material properties and to detect battery internal structural features, thus facilitating the identification of potential failures (see Fig. 2a). Kong et al. demonstrated that despite capacity and impedance measurements being in line with the manufacturers' specifications for commercial cells, CT images revealed quality concerns, such as electrode misalignment and gaps in the winding structure (Fig. 2b)[22]. Additionally, Wu et al. revealed that various defects could be observed in 18650 and pouch cells, using CT, such as welding burrs, overlapping of electrode tabs, and deflected electrodes[20]. As conventional electrochemical methods based on electrical testing of batteries are limited in terms of the failure information they can provide, CT can be utilized in partnership between industry and academia to assess the impact of new processes on assembly and packaging quality without disassembly[23].

*Battery wetting*

After packaging, the electrolyte infusion process in battery manufacturing comprises two phases: filling and wetting to introduce the appropriate amount of electrolyte into the cell, ensuring complete wetting of all pores in the cell materials. This ensures even current distribution. An excessive electrolyte-to-material ratio can reduce energy density, increase cell cost, and increase safety liability (as the electrolyte is a volatile substance)., Insufficient wetting leads to reduced capacity, rapid capacity fade, low power capability, and non-uniform solid electrolyte interface (SEI) layer growth during formation[24]. Despite the widespread recognition of potential time and cost savings for the industry, the academic (and patent) literature on electrolyte filling is currently insufficient to address the process in a reliable manner.

To assess battery wetting, electrochemical characterizations are commonly used. For example, high frequency resistance measured by EIS ) is directly related to the wetting state of the cell[25]. Additionally, OCV and Electrical contact resistance (ECR) measurement offer insights into the wetting state[26,27]. To catch information of non-homogeneous area imaging techniques using X-ray or Neutron sources have been used by researchers. Those study provide information on the effect of the cell design, electrode porosity and electrolyte and helped to optimize wetting process[28–30]. Neutron radiography, for instance, has revealed the impact of directional flow of electrolyte liquid within the electrode assembly, creating gas entrapments[31]. However, these techniques often require expensive equipment and may only address specific questions or issues. Moreover, X-ray imagine require contrast agent which hinder their utilization for in-line measurements[32]. A complement to previous methods, acoustic methods offer an inexpensive and high-quality visualization using sound waves. Acoustic sensing can then be passive[33] (measurements of the acoustic wave generated by the environment) or active[34] (by emitting sounds waves and analyzing their propagation, see Fig. 2c). Notably, ultrasound exhibit higher attenuation in electrodes or separators that are not adequately wetted, making acoustic sensing a preferred method for wetting monitoring[35]. Deng et al. employed an ultrasonic battery scanner to achieve 2D mapping of a commercial battery, enabling the visualization of electrolyte wetting, drying, and gas formation during cycling (Fig. 2d and e)[36]. This imaging technique enables rapid determination of the minimum required electrolyte injection volume and wetting time, facilitating the optimization of battery manufacturing processes.

*Battery formation cycles*

At the end of the conditioning, battery formation is one of the most crucial and most closely guarded processes in the manufacturing of batteries, particularly for lithium-ion batteries. It consists of the initial charging and discharging cycles that a battery undergoes before it is ready for use or sale. During this formation process, the electrodes (anode and cathode) and the electrolyte interact to create a solid electrolyte interphase (SEI) layer on the anode's surface. The formation protocols are intricate and require precise control of temperature, voltage, and charging rate to quickly produce uniform, electrically insulating, and facilitates ion conduction[37]. The SEI plays a pivotal role in battery performance as it acts as a protective barrier, preventing further reaction between the anode and electrolyte, thus reducing capacity fade and enhancing the battery's overall capacity, efficiency, and cycle life[38,39]. The complexity of SEI formation mechanism is the biggest challenge for the development of new formation protocol that is shorter, optimized, more reliable, and suitable for different active materials and electrolytes.

In industrial practice, electrochemical techniques are commonly used to monitor the cell's response during formation protocols, particularly by analyzing the first EIS spectrum semicircle, ascribed to the SEI layer's impedance and allows tracking of interface growth. To gain deeper insights into the mechanism behind battery formation, researchers directed their attention to observing the gas generated as a result of electrolyte decomposition. Several techniques have been employed to monitor pressure increases during cell formation cycles. Among them, electrochemical dilatometry and X-ray computed tomography enable non-destructive probing of gas generation[40]. For large-scale applications, operando pressure measurements with strain gauges mounted on cells are particularly suitable, allowing tracking of volumetric expansion during cycling. Louli et al. demonstrated the effectiveness of this technique in accurately monitoring SEI growth and effectively decoupling gassing from swelling[41]. Moreover, acoustic sensing allows to observe gas generation and SEI formation, thanks to the distinct propagation of acoustic waves depending on the materials' gas, porosity, and mechanical properties. Using acoustic transducer, Bommier et al. proved the ability of acoustic measurements to track operando gas generated during formation as well as silicon passivation (Fig. 2c)[42].

Recently, researchers have drawn inspiration from the photonic field and successfully integrated optical fiber sensors into batteries for thermal and mechanical monitoring[12]. Among them, Fiber Bragg Gratings (FBGs) sensors use periodic changes in an optical fiber's refractive index to reflect specific wavelengths of light, allowing them to measure various environmental factors such as temperature, strain, and pressure. The sensitivities to this factor can be decoupled through careful design of implantation, packaging, or sensor combinations. For example, by using microstructured optical fiber, that under its air-hole pattern design, is highly sensitive to hydraulic pressure, Huang et al. demonstrated the coincidence of pressure increase with thermal events measured by FBG sensors during the first charge of 18650 cells that are ascribed to electrolyte decomposition. From the FBG sensors measurements, heat was also calculated, and the team demonstrated that the cascade reactions leading to SEI could be observed with this technique[43].

However, as valuable as these approaches are, they cannot capture molecular insights occurring in the battery and the cascade of chemical reactions underlying the formation of the SEI. To close this gap, tilted FBGs were employed to measure the refractive index variation in the electrolyte during the formation cycles. Large refractive index variations were attributed to unstable electrolytes, providing a preliminary understanding of their stability but needing more detailed information about the underlying mechanism causing these variations[44]. To address these limitations, Miele et al. utilized photonic crystal fibers, commonly employed in developing biosensors for molecular recognition, in combination with microfluidics to monitor the evolution of an electrolyte by Raman spectroscopy during cycling in battery systems. However, the reported approach relies on an external pump to extract small amounts of electrolyte, limiting continuous monitoring and the ability to track changes in the electrode materials themselves[45]. Recently, researchers integrated IR fiber evanescent wave spectroscopy (FEWS) in commercial 18650 Na-ion cells (Fig. 2f). Fiber evanescent wave spectroscopy utilizes the interaction between guided light in an optical fiber and the evanescent field extending beyond its core to analyze the properties of substances or molecules near the fiber's surface. Using IR transparent chalcogenide fiber, electrolyte stability could be assessed in operando with excellent sensitivity, identifying the nature of electrolyte decomposition products and ion solvation dynamics as a function of voltage and current (Fig. 2g). Furthermore, embedding the fiber directly in the material enables the tracking of Li uptake-removal processes in positive electrode materials using IR spectroscopy[46].

The introduction of non-destructive battery characterization methods has the potential to improve the quality control of battery manufacturing processes, facilitating the identification of defects at every stage. In addition, the knowledge gained from these techniques will enable the rapid evaluation of new electrolytes and additives, facilitating the development of more effective formation protocols, ultimately resulting in higher production yields, improved cell quality and reduced costs. We anticipate the rapid integration of acoustic monitoring, while for more long-term goals, facilitated by synergistic efforts from different fields, IR fiber or similar spectroscopic technologies could be integrated to optimize formation protocols.

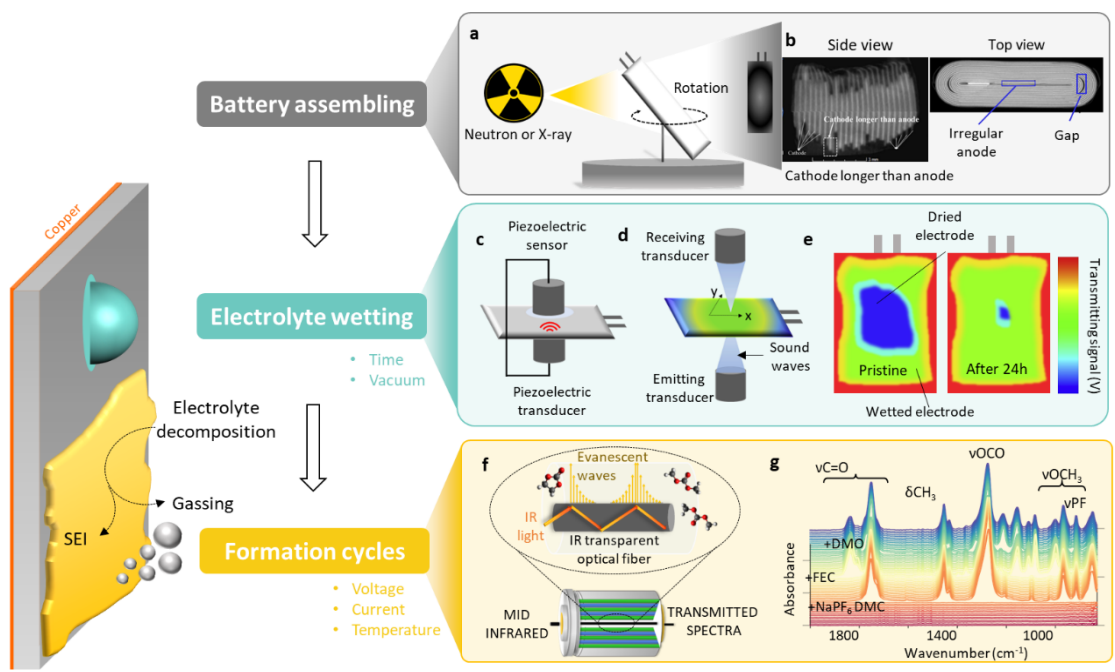

**Fig. 2 | Non-destructive characterizations to understand battery manufacturing process, including battery assembly, battery wetting and formation cycles. Battery assembling; a)** Schematic drawing of battery tomography characterization principle. **b)** CT cross-sections images taken from pouch cell after assembly, highlighting defaults in the welding (adapted from Ref.[22]). **Electrolyte wetting: c)** Schematic drawing of battery acoustic characterization. **d)** Schematic drawing of battery acoustic imaging characterization. **e)** Acoustic images of the wetting process (adapted from Ref.[36]). **Formation cycles; f)** Schematic drawing of infrared (IR) fiber evanescent wave (IR-FEWS) within chalcogenide glass fiber principle placed in 18650 cell. **g)** IR-FEWS absorbance spectra evolution during the addition of different electrolyte solutions (adapted from Ref.[46]).

II. During usage: Non-destructive characterization opportunities for electric vehicle.

EVs demand batteries with improved performance and safety. Currently, o be able to meet the vehicle's requirements for mileage and traction power, battery packs consist of hundreds or thousands of single cells in series and parallel connection (Fig. 3a). Among those individual cells, variation in capacities, voltages, internal resistances, and aging effects can lead to system failures. These inconsistencies, arising from factors like impacts, vibrations, temperature fluctuations, or internal defects from manufacturing, result in degraded battery performance and excess heat generation during operation. To tackle this, the battery management system (BMS) monitors and regulates the charging, discharging, temperature, and overall health to ensure safe and efficient battery operation. The current BMS relies solely on imprecise measurements of current, voltage, and external temperature at the battery pack level, which cannot detect individual cell degradation (see Fig. 3a).

Significant efforts have been directed toward enhancing the BMS through modeling techniques for battery state and thermal evaluation and prediction[47], including incorporating machine learning approaches to enable more accurate prediction of battery behavior and health[48]. Additionally, recent advancements in on board and online impedance spectroscopy have raised great expectations, as the impedance spectrum of a battery cell at a given instant depends not only on the state of charge (SOC), but also on many physical/chemical properties of the cell, such as aging, and chemical degradation effects field[49,50]. However, it's essential to recognize that impedance is defined solely for time-invariant systems. Consequently, it cannot identify transient phenomena that could elucidate the causes behind battery degradation.

Notably, mechanical stress and temperature variations can serve as early indicators of chemical or structural defects that could impact battery safety. Consequently, the parameters have to be carefully monitored in order to improve BMS and overall safety. Herein described below are sensors that hold the potential to monitor mechanical and thermal evaluations of cells within the battery pack.

*Thermal management*

Temperature is one of the main causes and consequences of battery degradation. High temperatures accelerate the electrolyte decomposition and SEI growth, leading to increased internal resistance, capacity degradation, and higher heat generation. Conversely, lower temperatures can induce lithium metal deposition, which can cause internal short circuits. Accurate temperature monitoring is then essential to detect such occurrences and protect the battery. Present BMSs commonly integrate thermocouple sensors onto the battery pack or specific cells to monitor surface temperature (Fig. 3b). These sensors are easy to use, compact and robust[51]. However, they exhibit relatively diminished accuracy and sensitivity compared to alternative temperature measurement techniques. For instance, micro temperature sensors can deliver enhanced precision and sensitivity temperature measurements within a confined area (see Fig. 3c)[52]. Nevertheless, these microsensors are more fragile, susceptible to damage due to their size, and more prone to electrical noise interference when compared to thermocouples.

FBG sensors provide measurement of temperature in real-time with a sensitivity of 0.1 °C, resistance to electromagnetic interference, possess a small size (around 100 µm), and are suitable for challenging environments due to their dielectric characteristics. Furthermore, the integration of multiple FBG sensors into a single optical fiber, where each sensor reflects light at a unique wavelength, facilitates the potential monitoring of separate cells and positions within a battery pack (Fig. 3d). However, surface measurements can't precisely catch the temperature fluctuations within the cell. Indeed, studies utilizing FBG sensors revealed that temperature discrepancies between the cell interior and surface ranged from 0.2°C to 5.4°C for 18650 cells cycled at rates from C/10 to 10C[43]. Furthermore, variations of up to 2°C were evident within the cell (see Fig. 3e)[53]. Furthermore, thermal measurements at different locations within and around the cell quantify the heat production during operation. For instance, fluctuations in generated heat, measured by FBG sensors, exhibited remarkable sensitivity to variations in C-rates within operational protocols like the Worldwide Harmonized Light Vehicles Test Procedure (WLTP, see Fig. 3f)[54]. Additionally, recent work has demonstrated the capability of optical fiber sensors to monitor the internal temperature and pressure of Li-ion 18650 cells prior to and during thermal runaway. This technique, which identifies safety early warnings through slope changes in temperature and pressure differential curves before thermal runaway, presents a pathway for enhanced accuracy in predicting and managing cell thermal behavior[55].

*Mechanical management*

Changes in electrode volume, electrolyte decomposition resulting in gas formation, or mechanical abuse to the battery pack can cause defects and thermal runaway. Consequently, mechanical cell management is as important as electrical and thermal management. Louli et al. demonstrated that employing a simple setup with a load cell sensor positioned between a cell and a plate enabled the measurement of reversible and irreversible volume expansion in Li-ion pouch cells (Fig. 3g). Notably, a direct relationship was established between irreversible pressure increase and performance degradation, along with polarization increase, confirming the need to monitor pressure [41]. The team also investigates the performance of cells containing SiO, Si-alloy, or carbon-coated nano Si electrodes and measured the substantial volume expansion encountered by Si-based electrode within pouch cells[56]. For large Li-ion pouch cell measurements, General Electric developed their own eddy current sensor (Fig. 3h). The device utilizes electromagnetic induction to measure changes in the size or shape of conductive objects, detecting expansions or contractions due to mechanical stress[57]. Additionally, by attaching FBG sensor at the surface of 18650 can, measurements of the battery SOC in real-time can be realized (Fig. 3i)[58]. Similar results were obtained with carbon nanotubes sensors attached at the surface of pouch bag (Fig. 3i)[59]. Diminishing the sensor dimensions permits mechanical measurements to be undertaken directly within the cell itself. By employing tin film strain gauges positioned at various locations, precise evaluations of the internal strain distribution within the cell can be studied (Fig. 3j)[60,61]. Recently, electrode scale measurements were achieved by optical fibers into electrodes (see Fig. 3k) for monitoring internal stress fluctuations, yielding insights into nano and micro silicon behavior within the electrode during cycling[62]. Lastly, it's important to note that optical fiber's versatility opens up a wide range of sensing opportunities; manipulating the fiber's microstructure can enhance sensitivity to pressure changes (Fig. 3l)[63]. Those fiber has previously been used for pressure measurement in 18650 batteries[43].

To enhance BMS, the integration of temperature and mechanical sensors, either internally or externally to the cell, offers an opportunity to enable early monitoring of degradation. These datas will enable battery manufacturers to take preventive measures on battery such as repair or replacement, reducing potential safety risks for users. Furthermore, these measurements will also contribute to extend batteries life by providing accurate assessments of their state-of-health.

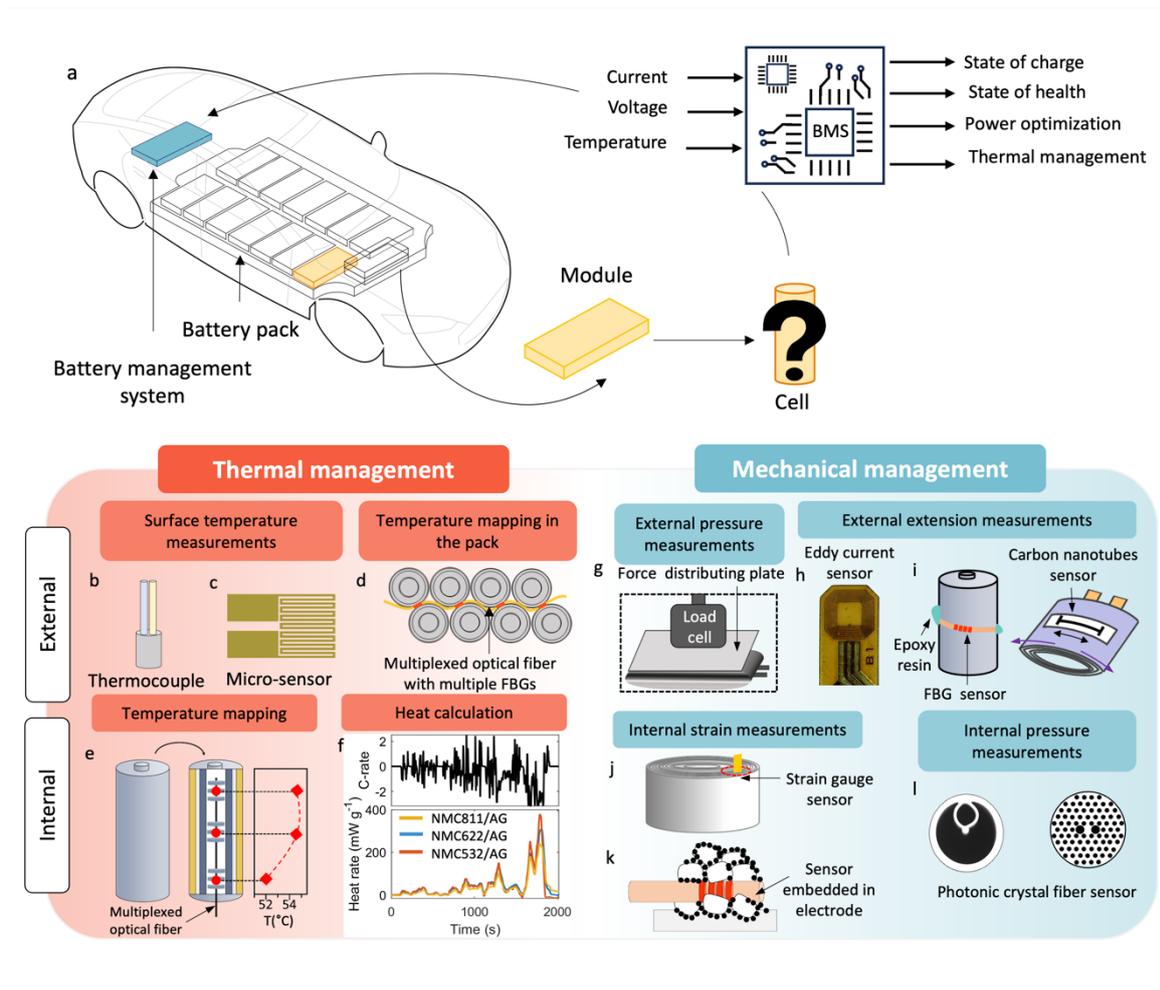

**Fig. 3 | Non-destructive characterizations opportunities for electric vehicle. a)** Challenge for battery characterization in electric vehicles. Currently, performance management is governed by the battery management system (BMS) which relies on a subset of temperature and voltage measurements at the module level only to estimate SOC, SOH and to optimize power and thermal management. **Thermal management; b)** and **c)** temperature sensors for cells external measurements[52]. **d)** Optical fiber with multiplexed FBG inscribed in it to measure cell temperature in module. **e)** Internal temperature mapping with FBG sensor (adapted from Ref.[53]). f) Heat calculated from FBG sensing during Worldwide Harmonized Light Vehicles Test Procedure (WLTP) of NMC-based cells[54]. **Mechanical management; g)** Load cell for external pressure measurements[41]. **j)** FBG sensor embedded in electrode for internal strain measurements[62]. **h)** Eddy current sensor [64]. **i)** External strain measurements with carbon nanotubes sensors (adapted from Ref.[59]) and FBG sensors[65]. **j)** Strain gauge sensor integrated in 18650 cell (adapted from Ref.[61]). **k)** FBG sensor embedded in electrode for internal strain measurements[62]. **l)** Photonic crystal fiber for pressure measurements inside cells [43,63].

### III. After usage: Characterizations of battery for end of life or second life option.

*Evaluation of retired battery for secondary application*

Recently, the electric vehicle (EV) sector has witnessed the emergence of a second-life market for batteries, encompassing retired EV batteries and marking the initial phase preceding the eventual recycling and recovery of materials. The battery pack is considered to have reached its end-of-life (EOL) when its capacity is below 80% of the nominal capacity within the domain of high energy-density electric vehicles. To take advantage of the remaining capacity, repurposing those batteries in less power-intensive applications, including energy storage systems and low-power devices has gained increasing interest. Especially, because based on data-driven battery degradation models with anticipated cycling conditions, it is considered that batteries commencing their second life with 70%–80% of their initial capacity can reasonably last more than 10 years in new applications[66].

However, degradation phenomena during battery aging are poorly understood and can impact battery safety. For instance, under certain cycling conditions, lithium ions can be reduced to lithium metal plating on the graphite anode's surface, forming uneven and dendritic lithium deposits that may cause internal short circuits and safety hazards. Additionally, other phenomena such as gas generation, cathode cracking, electrode delamination, binder decomposition, and current collector corrosion also impact battery aging and safety. Adding to the intricacy of the situation, cells within the same battery pack frequently demonstrate performance disparities, the worst cell in a series connected string determining the EOL. As a result, accurate evaluation of battery degradation at the cell level and determination of their true EOL status within second-life applications is essential. As part of this, battery packs have to be disassembled in order to '"screened, sorted and classified" the cells (Fig. 4a)[67]. These steps aim to identify cells that do not meet the requirement for second-life applications and to regroup batteries with comparable degradation levels and similar electrochemical behaviors[68].

*Electrochemical performance evaluation*

To quantify battery degradation most of the time electrochemical tests will be conducted, including open circuit voltage (OCV), internal resistance and capacity measurements. Among them, the incremental capacity (DV-IC) analysis developed by Dahn's group, can be used to assess the health and performance of a battery. Additionally, EIS spectroscopy can be used to measure battery's Ohmic resistance, charge transfer resistance, diffusion, electrode degradation, and state of health (SoH)[69]. However, interpreting the EIS data requires complex algorithms, either model based (using equivalent circuits) or data-driven based[70]. However, the intricate nature of battery degradation and the diversity of materials and electrolytes within battery systems limit the applicability of an universal model to understand battery degradation, leading to misinformation about battery aging[71]. Finally, as previously mentioned, EIS and electrochemical tests in general, gives averaged properties of the entire cell, potentially hindering safety risks arising from local defects. Especially during the cycling of large-size batteries, variations of voltage, current density, and local SoC over the battery can lead to non-uniform local heat generation under different operating conditions, resulting in non-uniform battery aging and increased local resistance over the battery surface as well as localized hot-spots and severe temperature gradients.

*Imaging techniques*

Advanced non-destructive and non-invasive techniques are required to characterize the battery health. Electrical or fiber-based sensors that require to be attached individually on or within the cells are then excluded. Imaging techniques such as X-ray, IR thermography and acoustics can provide rapid evaluation of the physical and mechanical properties of the cells. Those techniques give information on various parameters such as temperature, electrolyte degradation/gassing issues, structure deformation, and active material degradation. X-ray imaging is an extremely useful technique to study battery degradation phenomena, allowing for example to observe damaged at the particle scale (Fig 4b top)[72]. Notably, Toby Bond et al. applied synchrotron-based computed tomography (CT) techniques for imaging the internal morphology deformation of the $LiCoO_2$/graphite jelly-roll pouch cell due to the gas evolution under abuse conditions, as shown in the bottom schematic in Fig 4b. The study demonstrate the non-uniform expansion of the pouch cell jellyroll, with the most significant expansion observed in regions where the jellyroll had been deformed prior to abuse, suggesting that the stack pressure in these distorted areas is lower compared to the rest of the assembly[73]. Non-destructive measurements employing X-ray and neutron techniques are powerful instruments for understanding battery degradation at different scales. However, the prohibitive costs and extensive length of these experiments hinder their widespread industrial application.

IR thermography is a technique that can capture and visualize spatially non-uniform temperature variations in objects or surfaces using infrared radiation. Interestingly, IR thermography of cells during cycling shows that the cell's hottest spot initially lies near the positive tab but as discharge progresses becomes centered, findings consistent across 18650 and pouch cells (as indicated in Fig. 4c)[74,75]. Additionally, Zoran Milojevic et al. applied IR to demonstrate the orientation effect due to gravity having a significant effect on large-pouch cell aging and thermal behavior. The cells aged in flat orientation retained higher capacity than those aged in rotated orientation, which guides that the same orientation cells should be packed together for the secondary application to prevent the non-uniform aging behavior[76]. Precise IR thermography measurements can require costly IR camera and control of the environments, but the technique offers non-contact, rapid, and versatile temperature measurements, making it a valuable option for battery EOL characterization.

Acoustic measurement offers non-destructive, high-resolution, and real-time inspection capabilities. Deng et al. utilized acoustic imaging to evaluate electrolyte consumption and gas formation in NMC 532/graphite pouch cell. The ultrasonic imaging reveals that the cells are no longer well-wetted after 2 years of testing at 55°C, which demonstrated the real EOL of the battery[36]. Additionally, Chang et al. realized spatially resolved operando acoustic scanning of commercial pouch cells, the scanning time being 2 minutes only. With this custom design they observed the rate of electrolyte consumption during cycling along with spatial variations in interfacial roughening (Fig. 4d, left).

*Metallic Lithium diagnostic*

Besides the above parameters, diagnosing the amount of metallic Li in the battery is crucial for the safety of used batteries. In commercial LIBs, Li plating happens when the battery is cycled at high C-rate or low temperature. In a study by Ren et al., demonstrated that lithium plating was the key reason for the increased risk of battery thermal runaway after aging[77]. Indeed, nanosized Li metal and dendrites morphology formed upon aging can lead to internal short circuits, electrolyte decomposition, and exothermic reactions, leading to uncontrolled heat generation and escalating feedback mechanisms that intensify the reaction rates. Moreover, owing to its exceptionally high energy density, lithium has the potential to supplant the present graphite anode in the battery of the future. Hence, the precise detection of lithium plating is paramount in ensuring battery safety.

Interesting studies by Grey's group, used non-destructive in-situ nuclear magnetic resonance (NMR) to quantify the inactive metallic Li in $LiFePO_4$/anode-free cells and observed the corrosion behavior of the Li metal during storage[78]. Bingwen Hu et al. applied non-destructive electron paramagnetic resonance imaging (EPRI) to monitor the uniformity of Li plating and its influence on the evolution of the inactive metallic Li in $LiCoO_2$/anode-free pouch cells[79]. Acoustic characterization is a scalable, non-destructive and operando technique that can be utilized to detect Li metal plating in commercials LIBs. Bommier et al. demonstrate that acoustic measurements can be reliably used to evidence Li metal plating within commercial-scale LCO/graphite pouch cells during operation (Fig. 4d, right). Through this method, they establish a correlation between Li metal plating, current rate, and operational temperature and provide significant insights into the plating dynamics[42].

The successful screening and sorting of batteries for second life depends on our capability to understand the degradation mechanisms that interact with each other and simultaneously contribute to overall degradation. Achieving high quality control is essential to help the repurposing of batteries and ensure that they meet safety standards. Acoustic monitoring is a particularly promising technology for battery end of life assessment, providing information on wetting, gas generation and lithium plating. However, for widespread adoption of acoustic monitoring for EOL characterization the demonstration of quick tests that reliably assess the health of a battery with an unknown history.

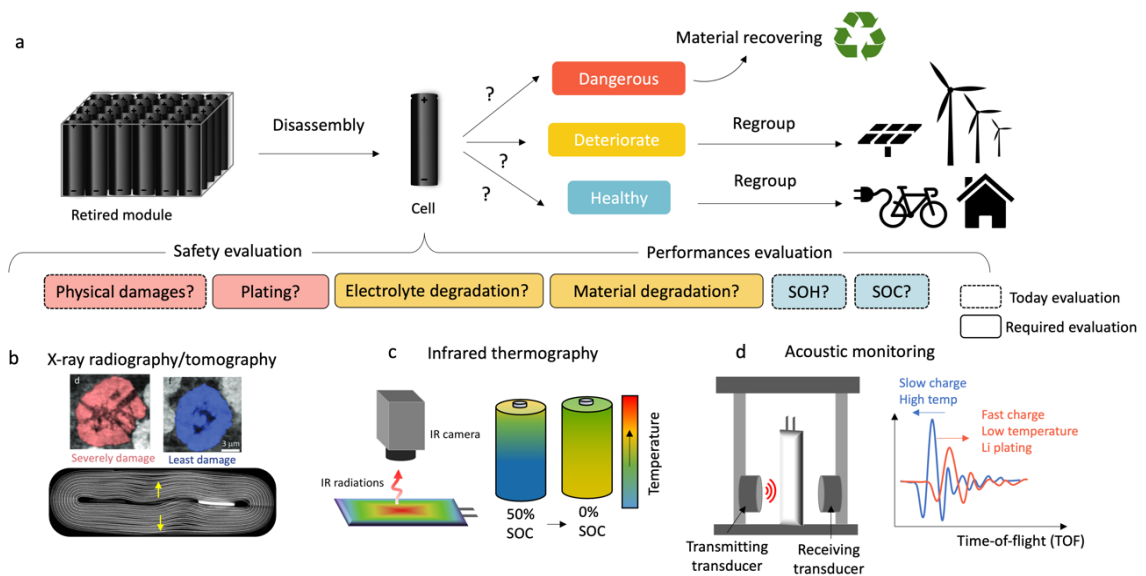

**Fig. 4 | Characterizations of battery for end of life or second life option. a)** Schematic steps for retired battery second life or end of life application. Battery pack and module are disassembled, screen and sort depending on their remining performances. Performances are evaluated using properties listed that are surrounded by dotted line. However, all the properties cited in the middle part of the figure should be evaluated to ensure maximal safety and performance of the second life application. **b)** Schematic drawing of the phenomena observed with X-ray imaging, including electrode aging[72] and cell swelling[73]. **c)** IR thermography principle and example of IR images of 18650 cells depending on their state of charge[75]. **d)** Acoustic scanning principle[80] and plating effect on time-of-flight signal[81].

## IV. Outlook

Throughout their lifecycle, batteries experience electrochemical and chemical side reactions such as SEI formation, electrolyte decomposition, lithium plating, dendrite formation, and physical structural changes including particle cracking, fragmentation, and delamination. Non-destructive characterization of batteries is essential to provide real-time monitoring of their behavior during usage, enabling optimization, safety, and longevity improvements while supporting second-life applications and recycling efforts. However, non-destructive battery characterization comes with its own set of challenges, including the need for cell-level monitoring, versatility, cost-effectiveness, and seamless integration into production lines or electric vehicles.

Acoustic monitoring emerges as a promising technique for non-destructive battery characterization due to its versatility, cost-effectiveness, and ability to assess critical battery properties such as wetting, SEI formation, and dead lithium, without compromising the battery's structural integrity. On the other hand, optical fibers are an ideal choice battery monitoring during usage, thanks to their compact size and high sensitivity to mechanical and thermal variations. However, successfully implementing these techniques will require collaborative efforts from various fields. For conditioning and end-of-life characterization the optimized ratio between experiment time and resolution will need to be found. Integrating sensors into electric vehicles requires defining the number, location, and positioning of these sensors around or inside the battery cell, while addressing the critical challenge of wiring and seamless integration into the system. Furthermore, careful consideration must be given to acquiring, storing, and processing sensor data through the Battery Management System (BMS).

For short term applications, non-destructive characterization plays a key role in refining digital twins. Battery digital twins are designed to replicate a physical battery's behavior and performance through real-time data and predictive modeling, enabling precise monitoring and optimization of battery systems. Integrating new data will refine those models allowing for precise behavior assessment. As machine learning grows in every domain, the convergence of this field with non-destructive characterization techniques presents a timely and promising opportunity to elevate our understanding and management of battery behavior and performance. The acquisition of realistic experimental data during battery operation has the potential to drive the development of precise machine learning and deep learning algorithms, further empowered by hardware advancements like significant GPU improvements, enabling accurate predictions of battery degradation and remaining lifespan.

Non-destructive characterization will also play a critical role in advancing the development and fostering the global adoption of the next generation of battery. For instance, lithium metal cells, being considered as the next-generation battery (whether liquid or solid), possess a lithium reservoir at the anode that hides performance degradation on capacity curves. Furthermore, the continuous formation of highly reactive dead lithium at the anode during cycling underscores the importance of developing new characterization techniques that can monitor the health of those cells[82,83]. In parallel, solid-state batteries present improved safety and higher energy density compared to conventional lithium-ion cells. However, the effects of stack pressure and chemomechanical volume changes within solid-state batteries can significantly impact cell performance[84,85]. Studies employing acoustic or optical sensing have already demonstrated the feasibility of characterizing the mechanical properties of these cells[86,62]. These studies as just the starting point and we anticipate that the evolution of the next-generation batteries will and must be accompanied by advancements in non-destructive characterization techniques.

In Europe, battery life cycle monitoring is already becoming a reality. Recent adoption of new regulations by the European Council aims to strengthen sustainability rules for batteries include the mandatory implementation of a battery passport for all electric vehicle (EV) batteries[87,88]. These measures have for objective to promote circular economy through end-of-life requirements, ensure battery improved and safe working conditions, while addressing fair competition and consumer information. These initiatives encourage collaboration between academia and industry to develop new non-destructive characterization techniques for commercial cells but also for new chemistry.

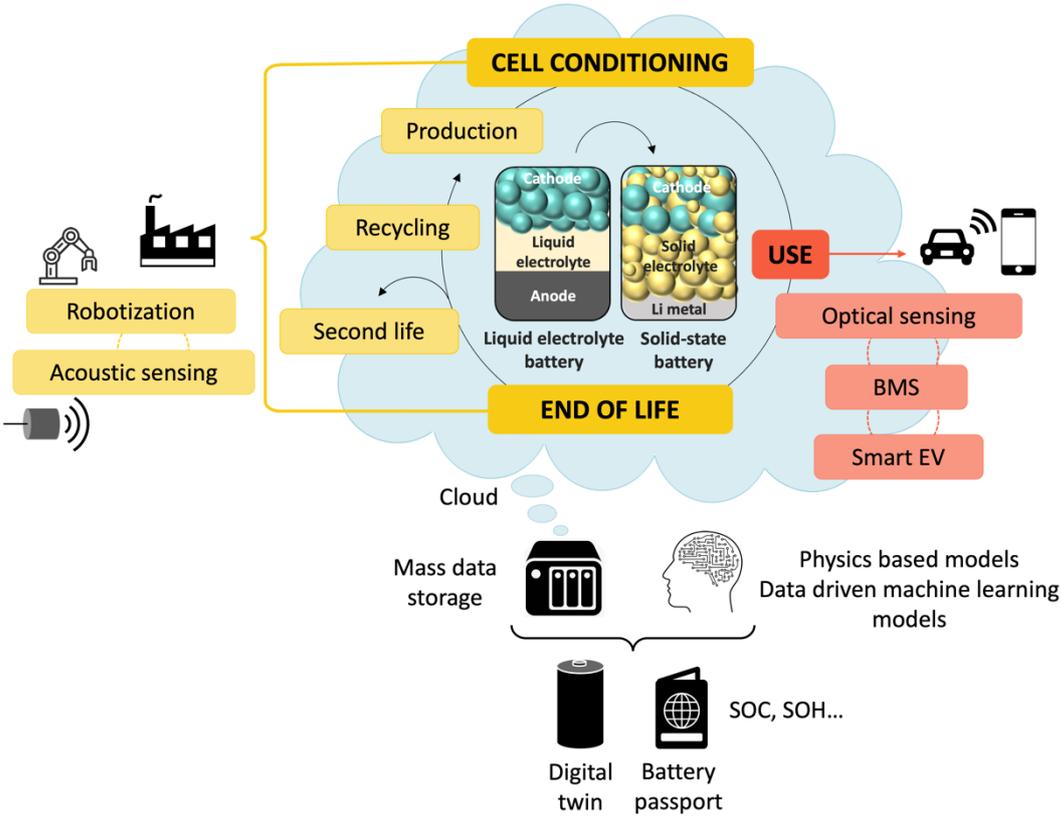

**Fig. 5 | Toward battery traceability.** Schematic drawing of the integration of non-destructive characterization for battery life cycle assessment. Acoustic and optical sensing techniques are suggested to image and measure degradation phenomena occurring throughout conditioning, usage, and end-of-life stages. The Integration of this characterization data into physics-based models or data-driven machine learning models can enhance battery digital twins, facilitating the adoption of battery passports.